\def\fnote#1{\footnote}

\documentclass[a4paper,12pt]{article}
\usepackage[dvips]{graphics}

\title{Active Stark Atomic Spectroscopy.}
\author{P.I.Melnikov\footnote{Corresponding author.
Address: Lavrentyev av. 11, BINP, 630090 Novosibirsk, Russia.
Phone: +7(383)2359 285. Fax: +7(383)235 2163. E-mail:
melnikov@inp.nsk.su}, J.B.Greenly, D.A.Hammer \\
{\small$^*$Budker Institute of Nuclear Physics,
630090 Novosibirsk, Russia} \\
{\small Laboratory of Plasma Studies, Cornell University, Ithaca, NY 14583}}
\date{ }
\begin{document}
\maketitle
\vspace{-1cm}
\begin{abstract}
Active Stark Atomic Spectroscopy (ASAS) method can be used to
determine the electric field in the diode of an ion or electron accelerator
as a function of position and time,  including the positions
of anode and cathode plasma emission surfaces (in order to obtain
the effective accelerating gap).  As possible probe beams, we suggest
the use of lithium and sodium atoms. The diagnostic provides a means
to measure diode quantities
spectroscopically with excellent spatial resolution.
\end{abstract}


\section{Introduction}
\par\medskip
\hspace{25pt}High-voltage vacuum diodes of various configurations have  been
used widely for generation of high-power electron and ion beams.
In these diodes, electric fields up to 10 MV/cm, magnetic fields of several
Tesla, and electron and ion current densities of kA/cm$^2$ are produced.
Dense, not fully ionized plasmas are generally produced at electrode
surfaces, either intentionally, as ion or electron sources, or unavoidably,
by explosive emission processes. The resulting dynamics of plasmas and
accelerated particles in these diodes require noninvasive temporally (ns)
and spatially ($<$mm) resolved diagnostics.
The single most criticle quantity for understanding of diode gap
processes is probably the electric field, but magnetic field, charge
particle orbits, and plasma motion incuding charge-exchange and
ionization of neutrals are also of great importance.

In typical high power ion diode configurations it is highly beneficial
to use spectroscopic methods for measurements.
Y.Maron {\it et al.} \cite{Maron2} used a spectroscopic technique
for the first direct measurements of the electric field distribution
in a  magnetically-insulated ion diode. They measured the Stark shift of
spectral lines of doubly-ionized aluminum ions as they crossed a diode gap.
This method can be
applied only to ion diodes with specially selected ion composition.
Moreover, its sensitivity is rather low due to the characteristics of
electron transitions in heavy ions. In particular, the precision of the
measurements in the above-cited work was 0.4 MV/cm.

Recently, similar experiments were carried out in the ion diode of the
PBFA-II accelerator \cite{Filuk}. The Stark shift of the {\em 3p}
level of lithium in electric field up to 10 MV/cm was determined in
these measurements
without resonance laser excitation of  atoms,  but with self-injection of
probe "charge-exchange" atoms from the partially ionized anode plasma
layer  into the diode gap. The probe atoms were produced because of the
absence of a dense plasma layer with zero electic field near the anode in
these experimets. They provide the first detailed investigation of  ion
diode acceleration gap physics in the high power pulses, and the first
observations of Stark shifts in a 10 MV/cm field.

Though the diagnostic methods  used for electric field measurements in ion
diodes so far have been successful, they are not generally applicable.
They cannot be applied, for example, to a proton-beam accelerator, nor
could they be used in a lithium-beam accelerator with a dense,
fully-ionized anode plasma. Furthermore, both of this techniques can
provide measurements only along a line of sight,  not at a ``point''.
In Ref. \cite{SovPhys} a method for measurement of the electric field in
diodes using probe atoms injected into the gap and excited by resonant
laser radiation was described. In this technique, called Active Stark
Atomic Spectroscopy (ASAS), the  Stark splitting  of a probe-atom spectral
line enables a calculation of the electric field with high time and space
resolution. Since the probe atom density is less than the density of the
background gas, this technique would not disturb the diode. However,
high sensitivity is provided by using resonant laser excitation to
saturate the population of the upper level of transitions of interest.
Because one can easily distinguish signal from noise by  simply omitting
the probe beam or tuning lasers away from resonance with transitions,
a reliable measurement can be made.

In Ref. \cite {SovTechPhys} measurements of the electric field in the
6-cm diode gap of the U-1 electron-beam accelerator by the ASAS
technique were  briefly described. A lithium atomic probe beam was injected
into
the gap before the high-votage was applied. Lithium levels with a
principal quantum number $n=4$ were excited through cascade transitions
using two dye-lasers. The bandwidth of the second laser was sufficiently
wide to excite the split components of interest. Spontaneous emission was
recorded with   1 mm spatial resolution by a monochromator combined with a
fiber-optic or electron-optic dissector (see \cite{SovTechPhys}).
These experiments enabled a direct measurement of the electric field
strength at the definite point as a function of time in the diode during
a {6-$\mu$s}, $\sim$1 MV voltage pulse. The electric field strength
measured in these experiments was 200--300 kV/cm, and the cathode and
anode emission surfaces were located as a function of time.

To apply the ASAS diagnostic method to ion-beam diodes, it is necessary
to take into account the specific ion diode conditions, including a
higher electric field strength and the presence of electron and ion
flows in the gap. In this paper, we consider  several experimental
arrangements for measurements of the electric field by  the ASAS
technique with lithium and sodium.

\section{\label{ASAS} Active Stark Atomic Spectroscopy in ion diodes}

\hspace{25pt}The selection of an atomic system, which can be used for Stark-splitting
measurements, depends upon: (i) the existance of  data on Stark-splitting
as a function of the electric field, (ii) the availability of suitable
lasers for a resonance transition, (iii) reasonable value of splitting
in the expected range of the electric field, (vi) a low rate of
field-ionization of the level. All these requirements are taken into
account in the present paper. Difficulties with the calculation of
Stark splitting in a strong electric field for atoms other than
hydrogen-like, and the requirement of acceptable resonance transitions,
suggest the alkali elements as probe atoms. Another attractive atom is boron,
because it can probably be excited with an intense KrF-laser. In this
section we will concentrate on Li and Na atoms only. We do not discuss in
this paper the problem of loss of the probe atoms by field or
impact ionization,
because this has already been analysed in detail in Ref. \cite{ SovPhys}.
In general, this is not a serious limitation.

\subsection{Stark splitting of lithium and sodium atomic levels}

\begin{figure}[t]
\centering
\unitlength = 1 cm
\resizebox{10 cm}{!}{\includegraphics{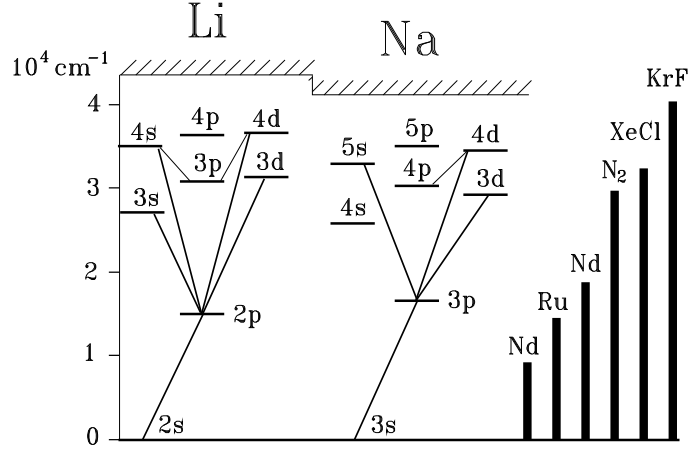}}
\caption[Grotrian diagrams for Na and Li.]
{Grotrian diagrams for Na and Li.}
\label{levels}
\end{figure}

\hspace{25pt}Partial Grotrian diagrams for lithium and sodium atoms are shown in Fig.\ref{levels},
in which the levels that can be used in the experiments are presented.
Motivation for selection of these levels will be given in the
next section. There is not much data on Stark splitting for atoms  in
intense electric fields.  Stark splitting for lithium at a field up to
500 kV/cm has been calculated in Ref \cite{SovPhys} for levels with
principal quantum number $n=4$. The {\em 4d} and {\em 4f} levels of
lithium differ only by $\Delta=$6.8 cm$^{-1}$. Because already for the
electric field of several tens kV/cm splitting becomes more then $\Delta$,
this leads to a rather strong splitting of the
{\em 4d}--{\em 4p} spectral line by
linear the Stark-effect\footnote{For the higher fields linear splitting
is distorced by influence of {\em 4p} level.}.

To calculate the Stark effect in the Li Schr\"{o}dinger equation
\begin{equation}
       \bigl(\hat{H}_0+\hat{V}\bigr)\Phi=E \Phi
\label{1}
\end{equation}
must be solved.
Here $\Phi$  is the wave function, $\hat{H}_0$ is the unperturbed
Hamiltonian, $E$ is the energy of the state,
$\hat{V}=e(\vec{r},\vec{F})$ is the perturbation Hamiltonian,
$e$ is the electron charge, $\vec{F}$
is the vector of  electric field strength.
When an electron of the {\em 2s} shell is excited to the Rydberg levels,
states are formed which are described by a wave function close to that of
the hydrogen atom.
The {\em 4p}, {\em 4d}, {\em 4f}  levels of lithium
are among such states, and
these are strongly
split in an electric field. Accordingly, we assume that the unpertubed  wave functions
of  {\em 4p$^m$}, {\em 4d$^m$},  {\em 4f$^m$} levels (which we designate as
$\Phi_1^m$, $\Phi_2^m$, $\Phi_3^m$, respectively) are close to
the classical hydrogen function
$\Psi_{4l}^m=R_{4l}\cdot Y_{4l}^m$. The superscript $m$ refers to
projection of the orbital angular momentum (quantum number $l$) on the $z$-axis.
For an electric field less then 500 kV/cm, the  perturbation intermixes only
nearby levels, {\em i.e.}, the solution of  Eq.(\ref{1}) is a
superposition of the wave functions $\Phi_i^m$.
If we choose the direction of the $z$ axis along the vector $\vec{F}$,
then the perturbation
$\hat{V}$ will not change the projection of the orbital momentum $m$ and
the desired wave function can be given the superscript $m$.

Let $\Phi^m=\Sigma c_i^m \Phi_i^m$. We multiply (\ref{1}) $\Phi_j^m$ and
integrate it,
taking into consideration the orthogonality of  $\Phi_i^m$
\begin{equation}
\sum_{i} (E_i+\hat{V}_{ji})c_i^m)=E^m \, c_j^m
\label{ce}
\end{equation}
Here, $E_{i}=<i|\hat{H}_0|i>$ are the eigenvalues of the energy of the
unperturbed
Schr{\"o}dinger equation, measured from the unshifted {\em 4d} level.
To carry out further
calculations, we must dertermine the $\hat{V}_{ji}^m$ values. It is not
hard to show that the replacement
of $\Phi_i^m$ by classical hydrogen wave functions $\Psi_{4l}^m$
gives an accuracy better
than 1$\%$ for the $\hat{V}_{ji}^m$ values. We obtain (see \cite{Bethe})
\begin{eqnarray}
  \hat{V}_{ii}^m=0\ , \ \   \hat{V}_{13}^m=\hat{V}_{31}^m=0 \ , \  \
  \hat{V}_{12}^m=\hat{V}_{21}^m=12\sqrt{(14-m^2)/5}\cdot ea_0 F \ ,\\
  \hat{V}_{13}^m=\hat{V}_{31}^m=6\sqrt{(9-m^2)/5}\cdot e a_0 F \ , \nonumber
\end{eqnarray}
where $a_0$ is the Bohr radius. With these values, we obtain the cubic
equation for eigenvalues $E_i^m$
\begin{eqnarray}
\bigl[E_{4l}^{\mid m\mid}\bigr]^3-\bigl(E_1+E_3\bigr)
\bigl[E_{4l}^{\mid m\mid}\bigr]^2-
\bigl(\bigl[V_{32}^{\mid m\mid}\bigr]^2+\bigl[V_{21}^{\mid m\mid}
\bigr]^2-E_1E_3\bigr)E_{4l}^{\mid m\mid}+
 \nonumber \\
+E_1\bigl[V_{32}^{\mid m\mid}\bigr]^2+E_3\bigl[V_{21}^{\mid m\mid}
\bigr]^2=0\ .
\end{eqnarray}
Here the absolute value $|m|$ is used due to the degeneracy of Stark
splitting with
respect to the sign of the orbital momentum projection. The
energy difference between {\em 4f} and {\em 4d} is $E_1$= 6.8 cm$^{-1}$
and the difference between {\em 4d} and {\em 4p} is $E_3$= 147 cm$^{-1}$
in the absence of an electric field. Solving the Eq'ns. (\ref{ce})
we find the
population of each split sublevel
\begin{equation}
n_{4l}^{\mid m\mid}=W_{\mid m\mid}\frac{\bigl(E_{4l}^{\mid m\mid}-E_1
\bigr)^2}{\bigl[V_{12}^{\mid m\mid}\bigr]^2+\biggl(1+
\frac{\bigl[V_{23}^{\mid m\mid}\bigr]^2}
{\bigl(E_{4l}^{\mid m\mid}-E_3\bigr)^2}\biggr)
\bigl(E_{4l}^{\mid m\mid}-E_1\bigr)^2}\ ,
\end{equation}
where $W_0$=1/5, and $W_1=W_2=2/5$.

The electric field mixes and splits
the {\em 4p}, {\em 4d}, {\em 4f} levels into eight $l, \mid m\mid$-sublevels:
$3,\mid m\mid$ ($\mid m\mid$=0,1,2)
(the 3,3-sublevel cannot interact with other sublevels since the orbital
moment projection is conserved by an electric field perturbation);
$2,\mid m\mid$ ($\mid m\mid$=0,1,2); and $1,\mid
m\mid$($\mid m\mid$=0,1). The index $l$ in this context
is formal and is connected only with the origin of a sublevel,
as long as  an orbital moment is not conserved in the electric field.
Stark splitting and split component intensities
for the above mentioned transitions of lithium are tabulated in
Ref.\cite{Preprint} for electric field up to 500 kV/cm.

In this paper we present more complicated calculations of Stark splitting of sodium lines
for electric field up to 5 MV/cm. The high strength causes strong interaction among the lines with
different principal quantum number $n$. Thus, for accurate calculations
the levels with low orbital quantum number $l$ must be involved. However, the
wave function of such levels are quite different from hydrogen wave
functions because
they have strong interaction with a non-hydrogen-like core. The
wave functions needed were obtained by solving the
hydrogen-like Schr{\"o}dinger equation, but using actual dependance
of energy levels of sodium from principal quantum number
$n$ as $(n+\Delta(l))^{-2}$, where $\Delta (l)$ is a quantum defect for
orbital quantum number $l$. These new wave functions were used in
the calculations of the Stark
effect in Na. The method of calculation was the same as described above for
lithium, but many times more levels were used
in the calculation process (levels with n up to 7, and {\em 8s} sublevel
was taken into account).

\begin{figure}[t]
\centering
\unitlength = 1 cm
\resizebox{14 cm}{!}{\includegraphics{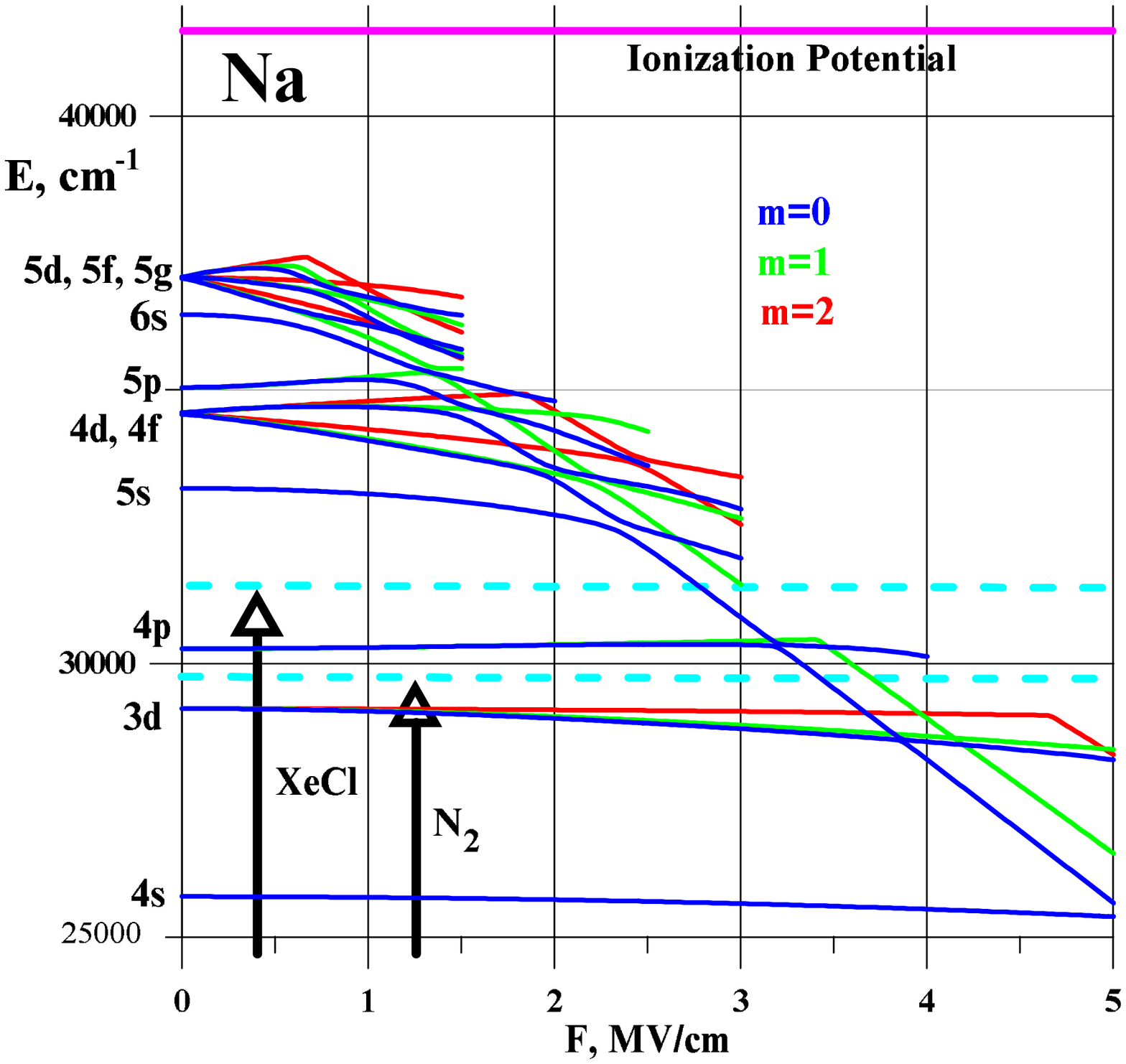}}
\caption[Overview of Stark splitting in Na.]
{Overview of Stark splitting in Na for m=0,1,2.}
\label{rnatr}
\end{figure}

The diagram of the splitting of the levels calculated with
STARK-I programme is shown on Fig.\ref{rnatr}.
These calculations
were performed for a wide range of levels with n=2,3,4,5 and  up to
5 MV/cm electric field strength. The Stark effect of higher-lying levels
is calculated up to 1.5 MV/cm, where these lines disappear
\cite{Geb,Trau,Lancz}. The lower level shifts are calculated up to
5 MV/cm. The main feature of the splitting seen in Fig.\ref{rnatr}
is the common
shift of all the levels to lower energy. This figure is a subject of
analysis for selection of working levels for experiments.

\section{Selection of the levels}

\begin{figure}[t]
\centering
\unitlength = 1 cm
\resizebox{8 cm}{!}{\includegraphics{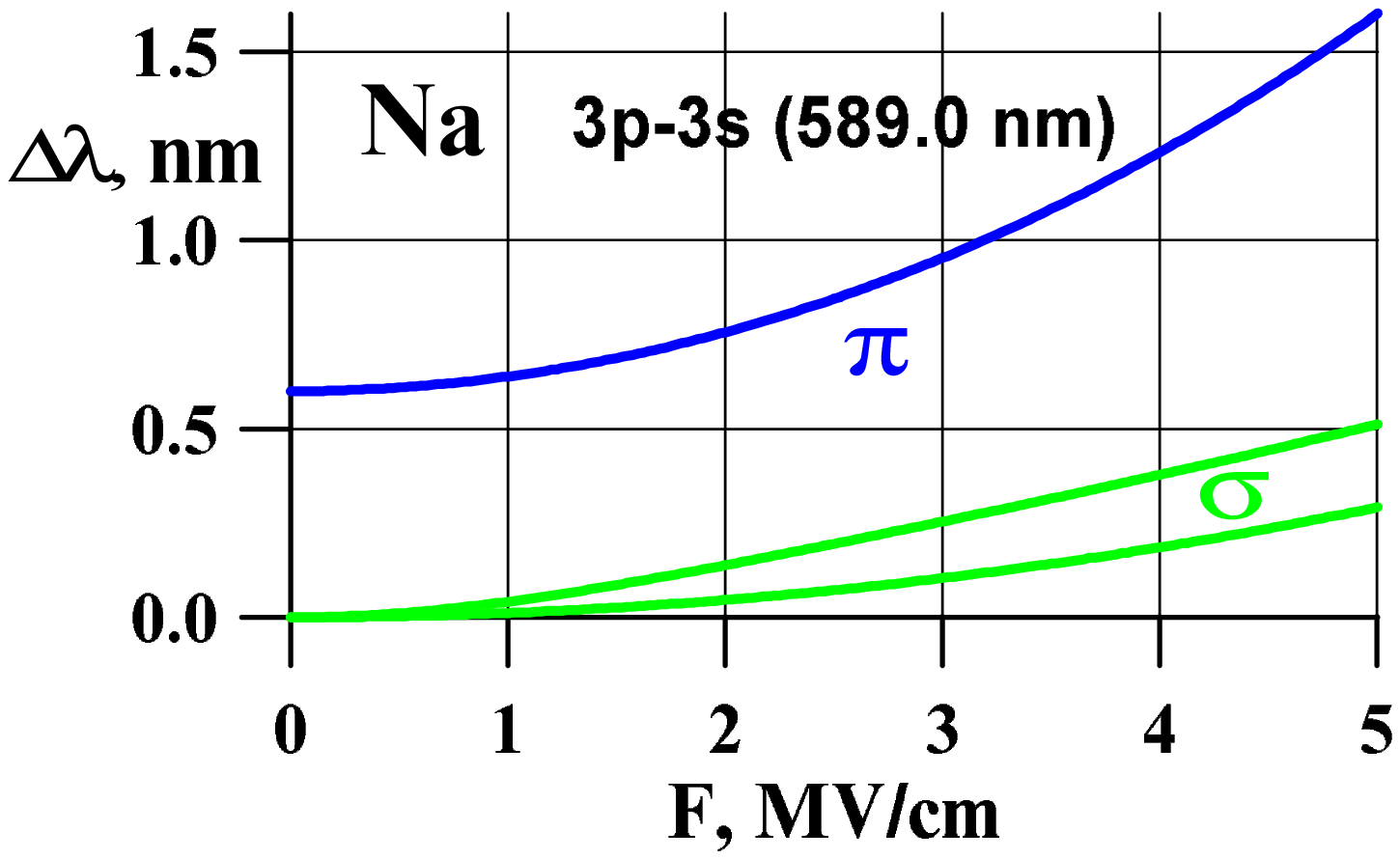}}
\caption[Splitting of {\em 3p-3s} transition.]
{Splitting of {\em 3p-3s} transition.}
\label{3p3s}
\end{figure}

\hspace{25pt}Preliminary consideration of the sodium levels shows us that only a few
levels meet all the requirements listed in Sec.\ref{ASAS}.
The first is the {\em 3p} level
that can be excited from the ground state level {\em 3s} by a dye laser with
wavelength 589.0--589.6 nm \cite{Atomic}. The {\em 3p}--{\em 3s} transition
is a strong one, having an Einstein coefficient $A=6.6\cdot 10^7$ s$^{-1}$;
therefore the lifetime of the {\em 3p} level is 15 ns. The second
possibility is the {\em 5s}
level, which can be  excited from the {\em 3p} level by a dye laser  with a
wavelength in the range 615.4--616.1 nm. The {\em 5s}--{\em 3p}
transition
is less strong then the previous one, with an Einstein coefficient
$A=0.6\cdot 10^7$ s$^{-1}$ (the lifetime is 150 ns).
The  {\em 4d} level can also be considered but it can be used for
measurements only with an electric field strength below 1.5 MV/cm,
because of field ionization.

Since the
lifetime of the {\em 3p} level is very short, this level must be excited
during the accelerator pulse. This is possible with a wide bandwidth
laser. Spontaneous fluorescence from the {\em 3p} level then gives the shift
of spectral line. The long lifetime of the {\em 5s} level
allows excitation before the voltage pulse, avoiding the need for wide laser
bandwidth.

\begin{figure}[t]
\centering
\unitlength = 1 cm
\resizebox{8 cm}{!}{\includegraphics{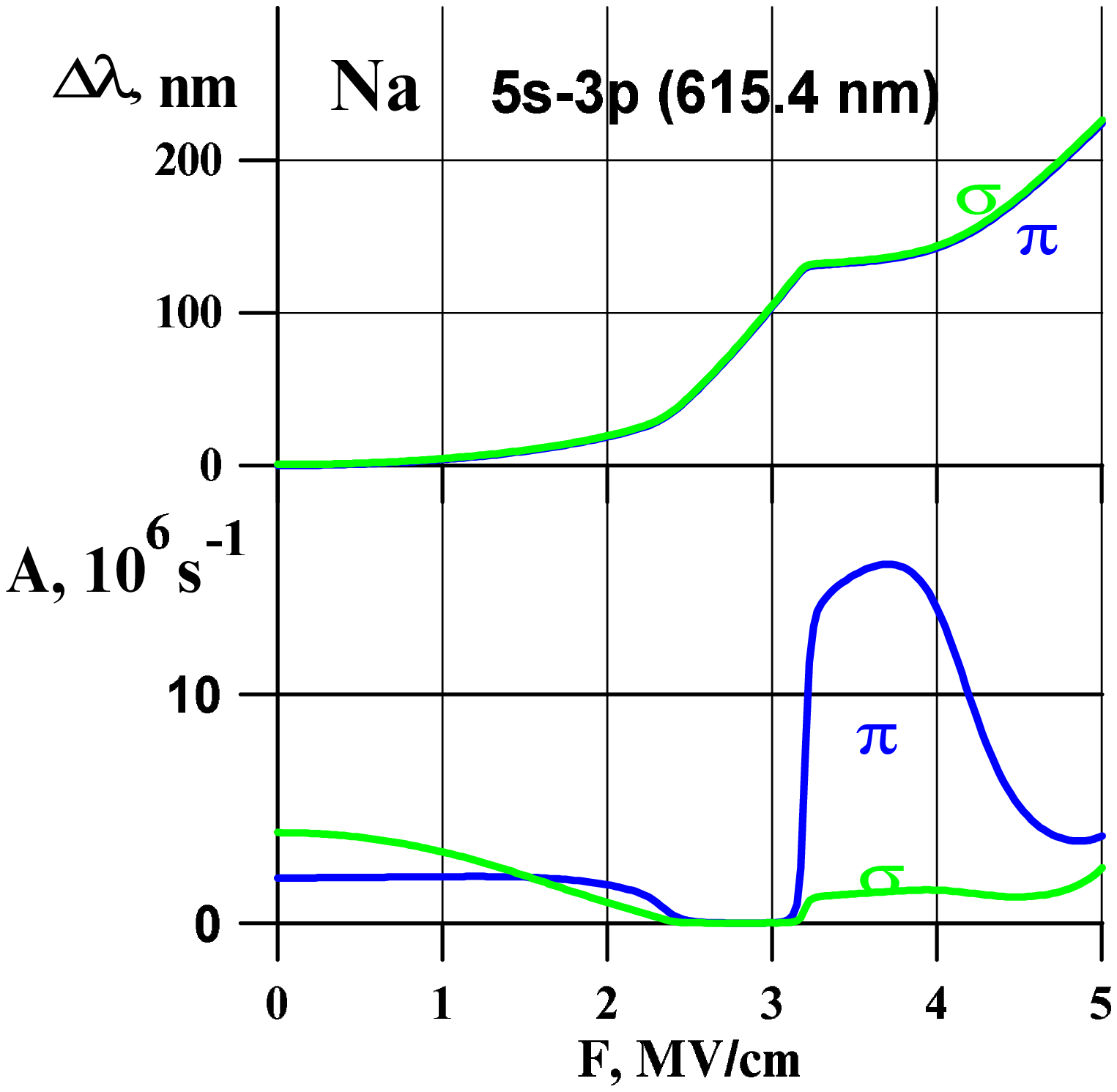}}
\caption[Splitting of {\em 5s-3p} transition.]
{Splitting of {\em 5s-3p} transition.}
\label{5s3p}
\end{figure}

The splitting of a fluorescence spectrum of the {\em 3p}--{\em 3s}
transition is illustrated in Fig.\ref{3p3s}. The splitting is calculated in terms of
wavelength shift from the initial wavelength 589.0 nm. The initial
splitting is caused by fine structure interaction.  Above 2 MV/cm
value of electric field strength the splitting becomes big enough for
measurement with common spectroscopy
apparatus. The intensities of lines are defined both by the population of
the upper level and  by dipole matrix element between upper and lower
states. If a high intensity laser is used the population of the  upper level
becomes saturated and does not depend on the electric field. The value of
population  is  $({g_2}/({g_1+g_2}))N=(3/4)N$ ($N$ is atomic density).
The dipole marix element (or the Einstein coefficient)
is almost not depend on
electric field
strength for this transition. It's value for $\pi$-component
is $2.2\cdot 10^{7}$ s$^{-1}$, for $\sigma$-component is $4.4\cdot 10^7$
s$^{-1}$. These calculated values are in a good agreement with the
Einstein coefficient from \cite{Atomic} as all the Einstein coefficients
mentioned below.

The level {\em 5s} has only one sublevel and the electric field shifts it.
However, this level has more  possibilities for transition then the
{\em 3p}.
In the high electric field all transitions are possible. In Fig.\ref{5s3p}, the only
one strong line acceptable wavelengthe close
to the visible range is shown. The intensities of lines are defined as in
previous case by the population of the level $({g_3}/({g_1+g_2+g_3}))N=N/5$
and the Einstein coefficient $A$. The
difference is that during the electric field pulse the population is not
supported by the laser and decreases with time due to fluorescence.
The Einstein coefficient of the {\em 5s}--{\em 3p} transition
changes dramatically with the electric field rise in contradiction to
{\em 3p}--{\em 3s} transition case.
The intensity in Fig.\ref{5s3p} is equal to the Einstein coefficient
$A$
when  F=0.

\begin{figure}[t]
\centering
\unitlength = 1 cm
\resizebox{8 cm}{!}{\includegraphics{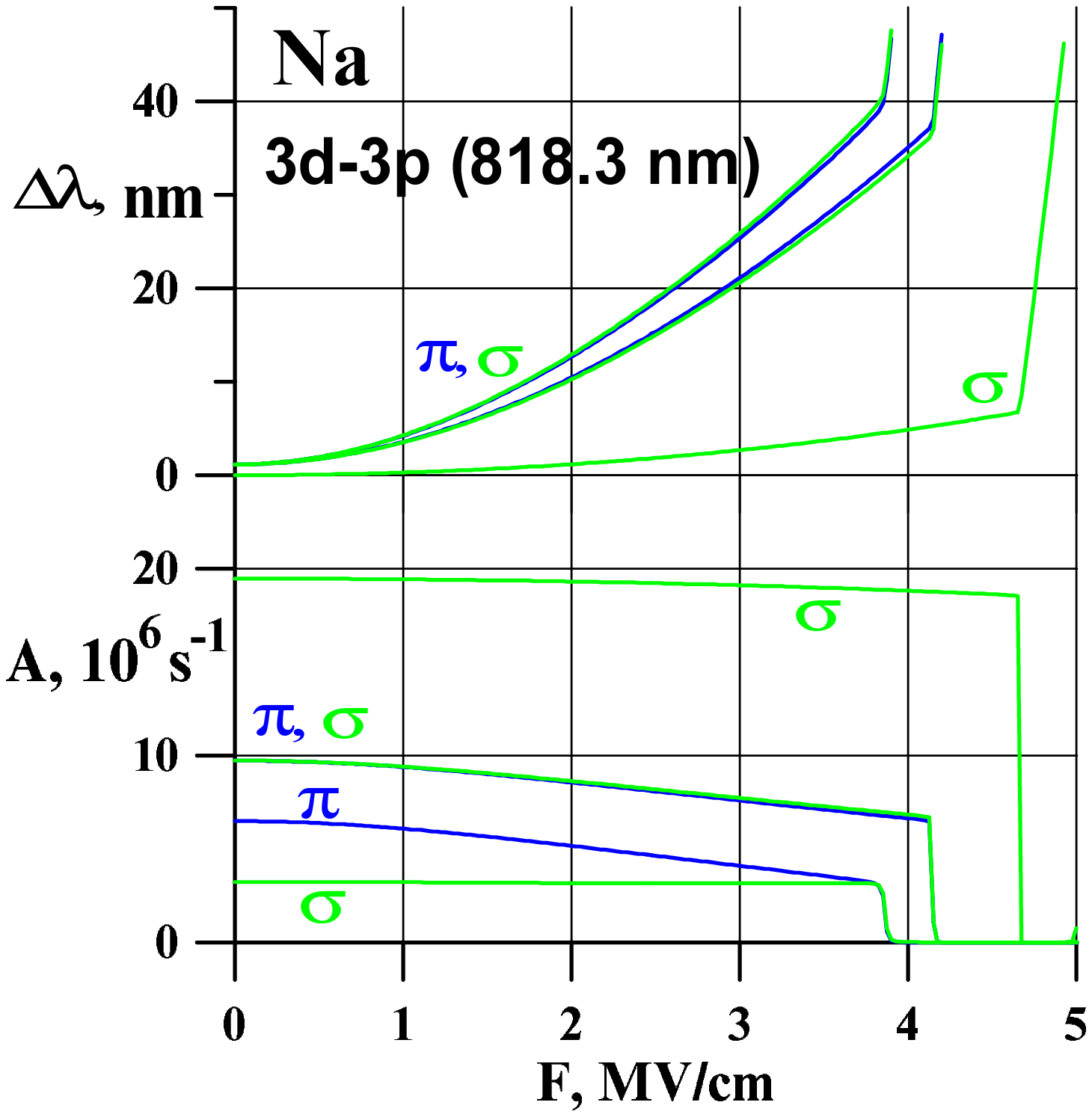}}
\caption[Splitting of {\em 3d--3p} transition.]
{Splitting of {\em 3d-3p} transition.}
\label{3d3p}
\end{figure}

 The {\em 3d} level can be used
with the same way as {\em 5s}. The splitting of the {\em 3d}--{\em 3p}
transition is shown in Fig.\ref{3d3p}.

Let us consider next a quite different possibility from the two
previous ones for the electric field mearsurement. If  we illuminate
the sodium
atoms with any laser, it could occur that levels can be excited in the
presence of the high
electric field during the pulse. For example, if we use a XeCl laser (308 nm),
the sodium atoms which are located where electric field strength is equal
to 2.7 or 3 MV/cm can be excited (see Fig.2).
An N$_2$ laser (337 nm)
will excite the atoms where the electric field is 3.4 or 3.7 MV/cm.
Any laser should excite the atoms at some value of electric field. If the
excited levels have a lifetime longer than the pulse duration the excited
levels could
provide spontaneous fluorescence and field measurement throughout the pulse.
We will call this way of exciting as "matching" (electric field causes the
atom's excitation energy to match the energy of the laser photon).

Li atoms can be used also in high electric field. However the first
transition {\em 2p}--{\em 2s} ($\lambda=670.8$ nm,
$A=4\cdot 10^7$ s$^{-1}$) seems to be difficult because of the small
splitting at fields less than 5 MV/cm. The {\em 4s-2p} ($\lambda=497.2$ nm,
$A=1\cdot 10^7$ s$^{-1}$), {\em 3d-2p} ($\lambda=610.4$ nm,
$A=7\cdot 10^7$ s$^{-1}$), and {\em 3s-2p} ($\lambda=812.6$ nm,
$A=0.2\cdot 10^7$ s$^{-1}$) transitions are preferable.
The first level has a
long lifetime, and  it can be used both with excitation just
before the pulse and in the "matching" mode. But in extremely high
fields ($>4$ MV/cm) the {\em 4s} level  will be ionized by the field.

\section{Electric field measurement in an ion diode gap}

\begin{figure}[t]
\centering
\unitlength = 1 cm
\resizebox{8 cm}{!}{\includegraphics{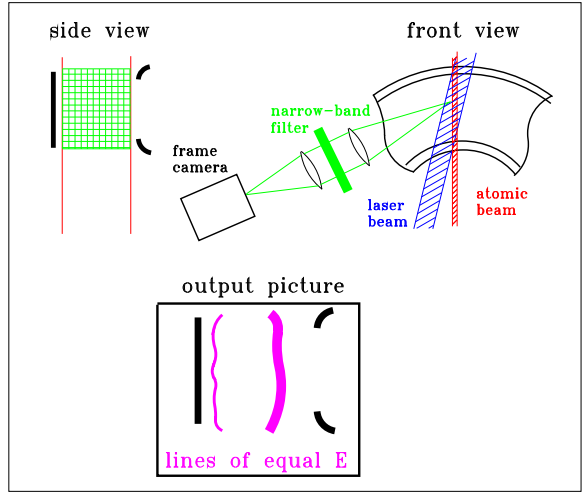}}
\caption[Scheme for observation of equal $|\vec{F}|$ lines.]
{Scheme for observation of equal $|\vec{F}|$ lines.}
\label{sch1}
\end{figure}

\hspace{25pt}There are a number of ways of observing the ASAS fluorescence.
Figure \ref{sch1} shows a way to observe lines of equal electric
field strength $|\vec{F}|$.
A thin rectangular (slab) beam  of thermal sodium atoms is introduced into
the ion gap. The wide dimension of the slab is parallel to the axis
so that the atoms stretch between anode and cathode. The atoms are
excited by slab  laser beams that have a small pitch angle to the
atomic beam in a plane parallel to the anode so that a wide area of
atoms will be excited and emit spontaneous fluorescence. An optical
system with a narrow band filter gathers the fluorescence and gives
an image of the fluorescing area on a frame camera. The filter selects
a narrow bandwidth (1$\div$10 nm) of the light
which corresponds to a definite electric field strength.
The resulting image will show light along lines of equal $|\vec{F}|$.

A similar concept can be realized with "matching" method. Instead of using
 a filter
the "matching" method of excitation would used. A short-pulse laser beam
irradiates the
atomic slab.  Atoms are excited only
in the regions with  definite electric field strengths, where the split
components of the spectrum are resonant with the laser wavelegth.
The spectral apparatus is not needed.
The resulting image shows the lines of constant $|\vec{F}|$. The time gate of the
frame camera can be wide because the moment of measurement is defined by
the laser pulse. This method appears to be more sensitive than the first.

If the long pulse duration laser will be used for excitation of
the levels with a short life time in the "matching" mode, it is possible to get the movies:
a set of pictures with lines of definite $|\vec{F}|$ in different
moments of time.

\begin{figure}[t]
\centering
\unitlength = 1 cm
\resizebox{16 cm}{!}{\includegraphics{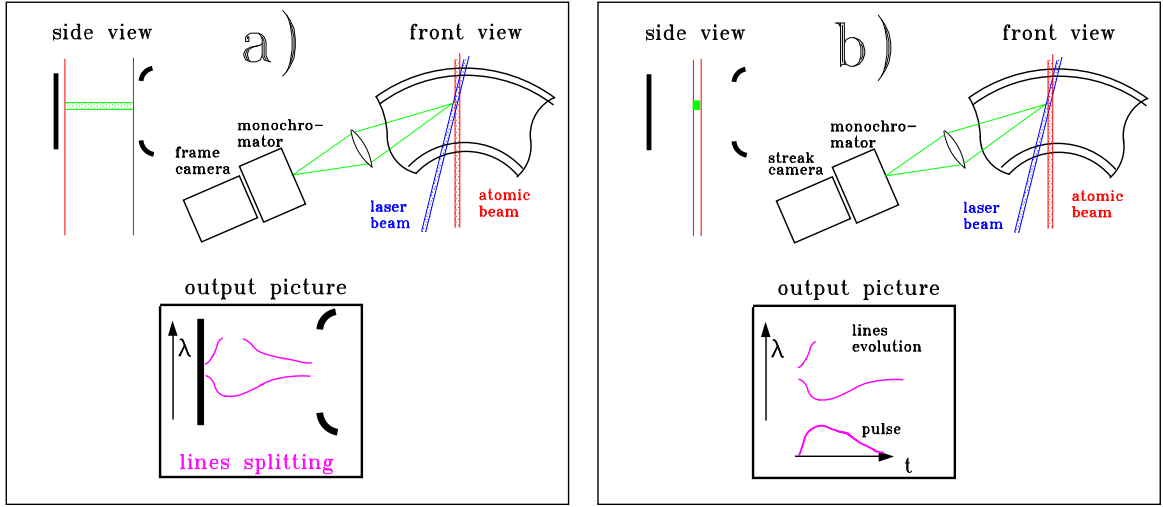}}
\caption[Schemes for electric field measurement.]
{Schemes for electric field measurement.}
\label{sch2}
\end{figure}

A method using a monochromator is illustrated by Fig.\ref{sch2}a. A thin coloumn of
excited atoms is imaged by a lens on the input slit of the monochromator so
that the distance along the slit images  the distance between anode and
cathode. The output of the frame camera gives a two-dimensional picture:
the horizontal axis is the distance between anode and cathode, and the
vertical axis is wavelength. The electric field variation in the gap can be
obtained from this picture. In addition the effect of line disappearance
provides a further check on the electric field measurement.

Replacement of the frame camera by a streak camera provides the possibility
to see the dependance of a local electric field strength in a time
(Fig.\ref{sch2}b).
The effect of field ionization can also be observed.

\section{Summary}

\hspace{25pt}The choice of concrete scheme of realization of ASAS technique
depends on the device  where the diagnostics whould be applied and
the goals of the experiment. But it must be pointed out
that the most simple way for getting the result is the "matching"
mode of ASAS. It is needed only one laser and unit for
registration. The spectral apparatus is not needed!

\end{document}